\documentclass[10pt,twocolumn,letterpaper]{article}

\usepackage{arxiv}
\newtoggle{usetimes}
\newtoggle{itabs}
\toggletrue{usetimes}  

\iftoggle{usetimes}{\usepackage{times}}{}
\usepackage{amsmath,amssymb,amsfonts,dsfont,pifont,bm,bbm,mathrsfs,mathtools,nicefrac}
\usepackage{algorithm,algpseudocode,listings}
\usepackage{booktabs,multirow,adjustbox,diagbox,threeparttable}
\usepackage[pagebackref=false,breaklinks=true,colorlinks=true,citecolor=blue,bookmarks=false]{hyperref}
\usepackage{cleveref}  
\usepackage{xcolor}
\usepackage{makecell}
\usepackage{stfloats}
\usepackage{setspace}
\usepackage{array}
\usepackage{xcolor}
\usepackage{cite}
\usepackage{enumitem}

\crefname{section}{Sec.}{Secs.}
\Crefname{section}{Section}{Sections}
\crefname{table}{Tab.}{Tabs.}
\Crefname{table}{Table}{Tables}
\crefname{figure}{Fig.}{Figs.}
\Crefname{figure}{Figure}{Figures}
\crefname{equation}{Eq.}{Eqs.}
\Crefname{equation}{Equation}{Equations}
\crefname{theorem}{Thm.}{Thms.}
\Crefname{theorem}{Theorem}{Theorems}
\crefname{algorithm}{Alg.}{Algs.}
\Crefname{algorithm}{Algorithm}{Algorithms}
\usepackage{listings}
\definecolor{codegreen}{rgb}{0,0.6,0}
\definecolor{codegray}{rgb}{0.5,0.5,0.5}
\definecolor{codepurple}{rgb}{0.58,0,0.82}
\definecolor{backcolour}{rgb}{1.0,1.0,1.0}
\lstdefinestyle{mystyle}{
    backgroundcolor=\color{backcolour},
    commentstyle=\color{codegreen},
    keywordstyle=\color{magenta},
    numberstyle=\tiny\color{codegray},
    stringstyle=\color{codepurple},
    basicstyle=\ttfamily\scriptsize,
    breakatwhitespace=false,
    breaklines=true,
    captionpos=b,
    keepspaces=true,
    numbers=left,
    numbersep=5pt,
    showspaces=false,
    showstringspaces=false,
    showtabs=false,
    tabsize=2
}
\begin{document}

\title{Enhancing Unsupervised Audio Representation Learning via \\
Adversarial Sample Generation}

\author{
    Yulin Pan$^{1}$ \quad
    Xiangteng He$^{2}$ \quad
    Biao Gong$^{1}$ \quad
    Yuxin Peng$^{2}$ \quad
    Yiliang Lv$^{1}$
    \\[5pt]
    $^1$Alibaba Group, Hangzhou, China \\
    $^2$Wangxuan Institute of Computer Technology, Peking University, Beijing, China \\
}

\maketitle

\begin{abstract}

Existing audio analysis methods generally first transform the audio stream to spectrogram, and then feed it into CNN for further analysis. A standard CNN recognizes specific visual patterns over feature map, then pools for high-level representation, which overlooks the positional information of recognized patterns.
However, unlike natural image, the semantic of an audio spectrogram is sensitive to positional change, as its vertical and horizontal axes indicate the frequency and temporal information of the audio, instead of naive rectangular coordinates. 
Thus, the insensitivity of CNN to positional change plays a negative role on audio spectrogram encoding.
%
%
To address this issue, this paper proposes a new self-supervised learning mechanism, which enhances the audio representation by first generating adversarial samples (\textit{i.e.}, negative samples), then driving CNN to distinguish the embeddings of negative pairs in the latent space.
%
%
Extensive experiments show that the proposed approach achieves best or competitive results on 9 downstream datasets compared with previous methods, which verifies its effectiveness on audio representation learning.

\end{abstract}
\section{Introduction}\label{sec:intro}

Audio analysis is a highly important technology with a wide range of applications, such as speech recognition \cite{gulati2020conformer, kriman2020quartznet}, acoustic scene classification \cite{ford2019deep,piczak2015environmental,mesaros2018multi}, music recognition \cite{snyder2015musan,engel2017neural} and so on. 
Encouraged by the great improvement in image analysis \cite{he2016deep,szegedy2015going,tan2019efficientnet}, the mainstream audio analysis methods \cite{kong2020panns,saeed2021contrastive,hershey2017cnn,gong2021psla} generally transform the audio into spectrogram firstly, and then obtain its semantic representation by CNN for further analysis. 
Limited to the scale of manually labelled data, recently self-supervised learning \cite{saeed2021contrastive,shor2020towards,tagliasacchi2020pre} is widely applied in audio analysis, to effectively learn better audio representation from the massive unlabelled data.
\emph{However, both CNN and existing self-supervised learning methods have limitations on audio representation learning.}

\begin{figure}[!t]
  \centering
  \includegraphics[width=\linewidth]{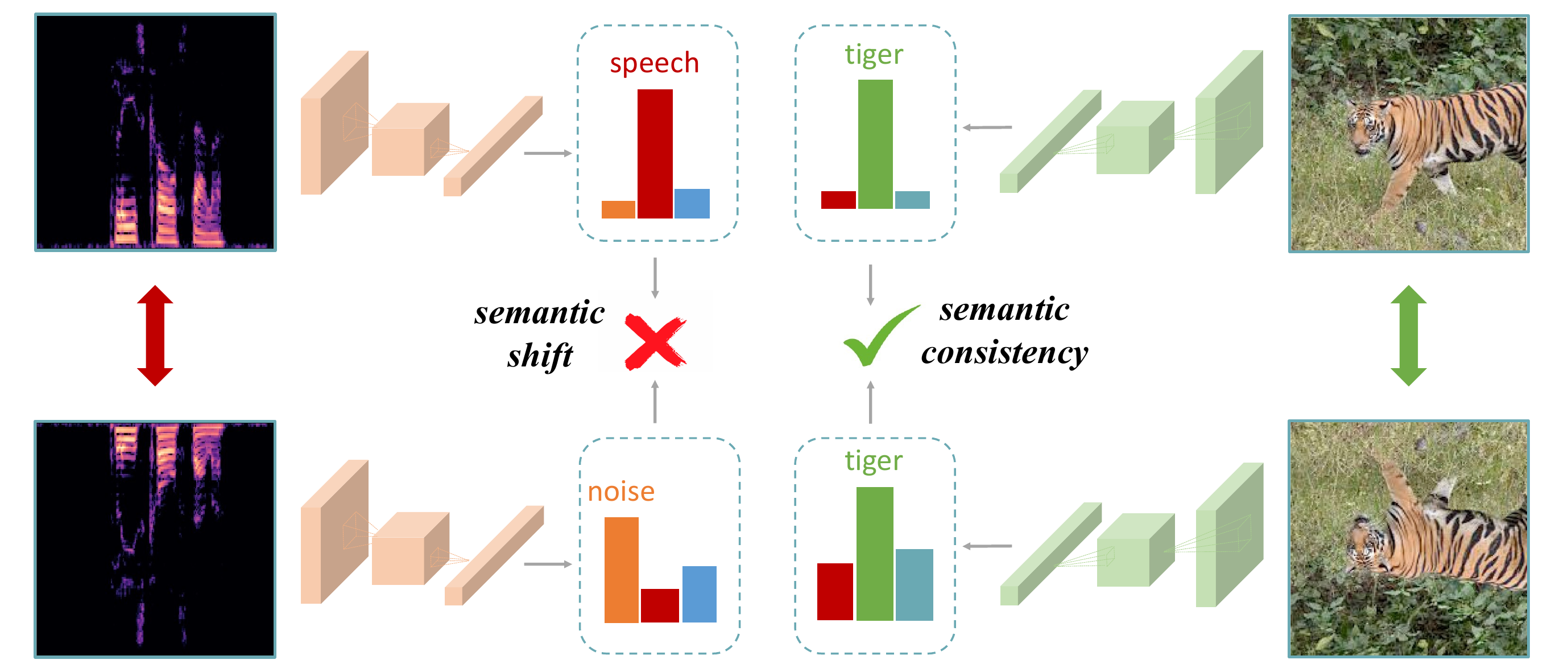}
  \caption{\textbf{The sensitivity to positional transformations of audio spectrogram (left) and natural image (right)}. The semantic of ``\textit{tiger}'' image stays consistent before and after it is flipped, whereas the semantic of audio spectrogram changes.}
  \label{fig:intro_1}
\end{figure}

We first introduce the audio spectrogram to discover the key issues involved.
The audio spectrogram is a two-dimensional matrix and can be seen as one-channel image. However, spectrogram should not be treated as a kind of natural image, because it shows some dramatically different characteristics. 
As illustrated in Figure~\ref{fig:intro_1}, the positional transformation of an audio spectrogram usually leads to semantic change, which is different from natural images.
Specifically, 
(1) the vertical axis of spectrogram represents the frequency of the audio, whose distribution reflects some inherent attributes of the acoustic source, and is the key characteristic in audio analysis. 
For example, increasing the voice pitch greatly while keeping the talk content, has a significant influence on speaker identification.
(2) The horizontal axis indicates the temporal information of the audio, where elements are arranged in time order. 
If we flip a spectrogram of speech audio along time dimension, the speech content changes thoroughly. 
To sum up, the audio spectrogram is sensitive to both frequency and temporal change, which we call \textbf{positional transformation}.

\begin{figure}[t]
  \centering
  \includegraphics[width=\linewidth]{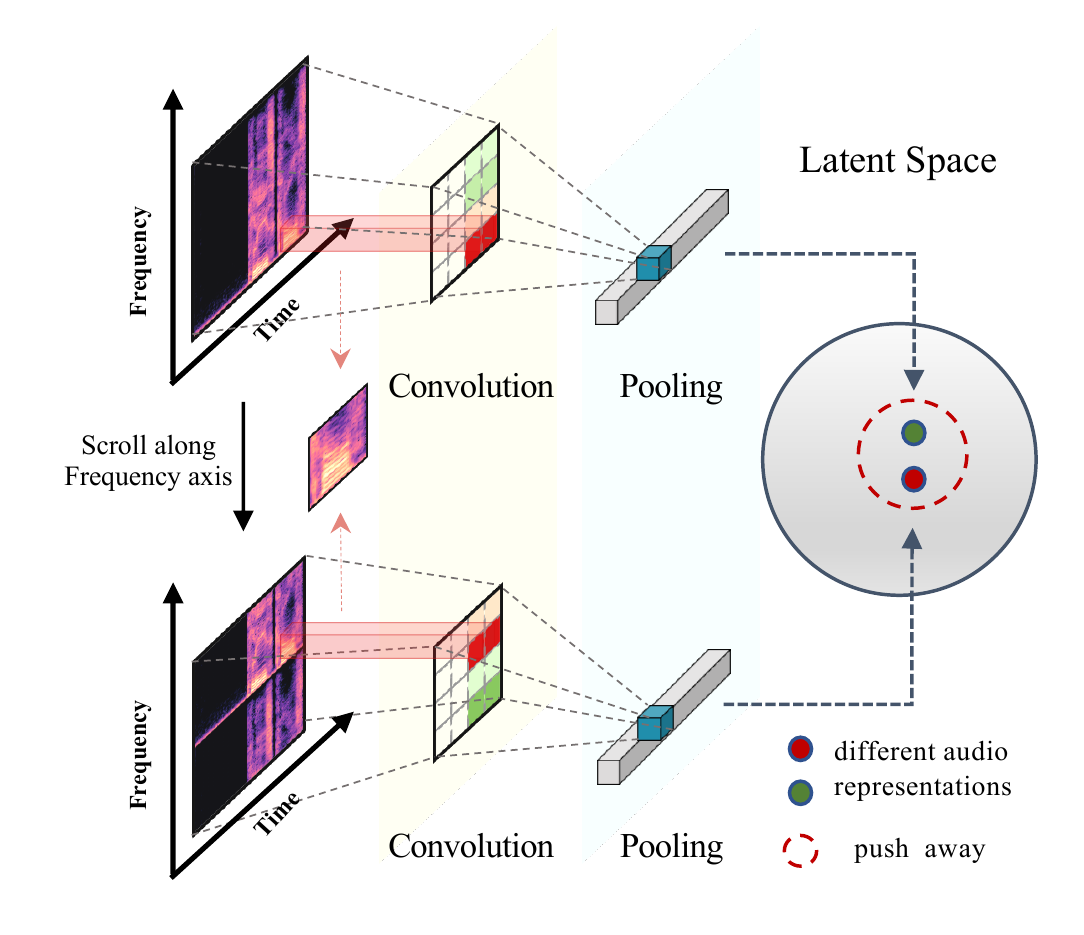}
  \caption{\textbf{Illustration for the positional insensitivity of CNN}. The two spectrograms which contain same visual patterns at different positions are irrelevant to each other. However, they are encoded as same embedding after pooling operation.}
  \label{fig:intro_2}
\end{figure}

However, CNN is insensitive to positional transformation.
%
In general, a fully convolutional network is a cascade of multiple Conv-Pool units and each unit is a combination of convolution layers and pooling layers. 
Such a cascade structure helps CNN progressively learn representation and output high-level features.
Inside Conv-Pool unit, CNN recognizes specific local visual patterns with convolution kernels over feature map, and then pools the feature map, overlooking the positional information. 
As depicted in \cref{fig:intro_2}, the two operations in Conv-Pool unit cause that the learned representation focuses on whether a specific visual pattern appears or not, while ignores its position. 
%
Since the positional information is essential to the semantics of audio spectrogram, we empirically consider the CNN has disadvantage on audio analysis.

In addition to this, existing self-supervised learning methods in audio field usually optimize a latent space by mapping semantic-relevant pairs (positive pairs) closer and pushing semantic-irrelevant pairs (negative pairs) away~\cite{saeed2021contrastive,fonseca2020unsupervised,jansen2018unsupervised}.
%
However, none of them consider about the positional transformations of audio spectrogram.

To address this issue, this paper introduces a new self-supervised learning framework based on Adversarial Sample Generation (ASG) to learn robust audio representation. The contributions can be summarized as follows:

\begin{itemize}[leftmargin=10pt]
\setlength{\itemsep}{0pt}
\setlength{\parsep}{0pt}
\setlength{\parskip}{0pt}
\item We introduce the \textbf{Adversarial Sample Generation} mechanism for audio contrastive learning to enhance the audio representation learning, based on the discovery of the sensitivity of audio spectrogram to positional transformations, which is different from image analysis but important to audio analysis.
Given a spectrogram, our ASG generates adversarial samples by translating or flipping it along time or frequency axis.
%
The generated adversarial examples share the same local visual patterns with original view, challenging CNN to pay attention to not only its appearance but also its position.
\vspace{3pt}
\item A tailored framework is introduced to learn better audio representation from a large scale unlabeled audio data, then transfer to the downstream tasks.
Following previous work, we evaluate the pre-trained representations on 9 downstream tasks under two different settings: \textit{(1) frozen}: training a single linear layer on the top of the pre-trained feature extractor; \textit{(2) fine-tuned}: end-to-end training the whole framework. 
The proposed ASG achieves state-of-the-art performance, which demonstrates its effectiveness.

\end{itemize}

\begin{figure*}[!ht]
  \centering
  \includegraphics[width=\linewidth]{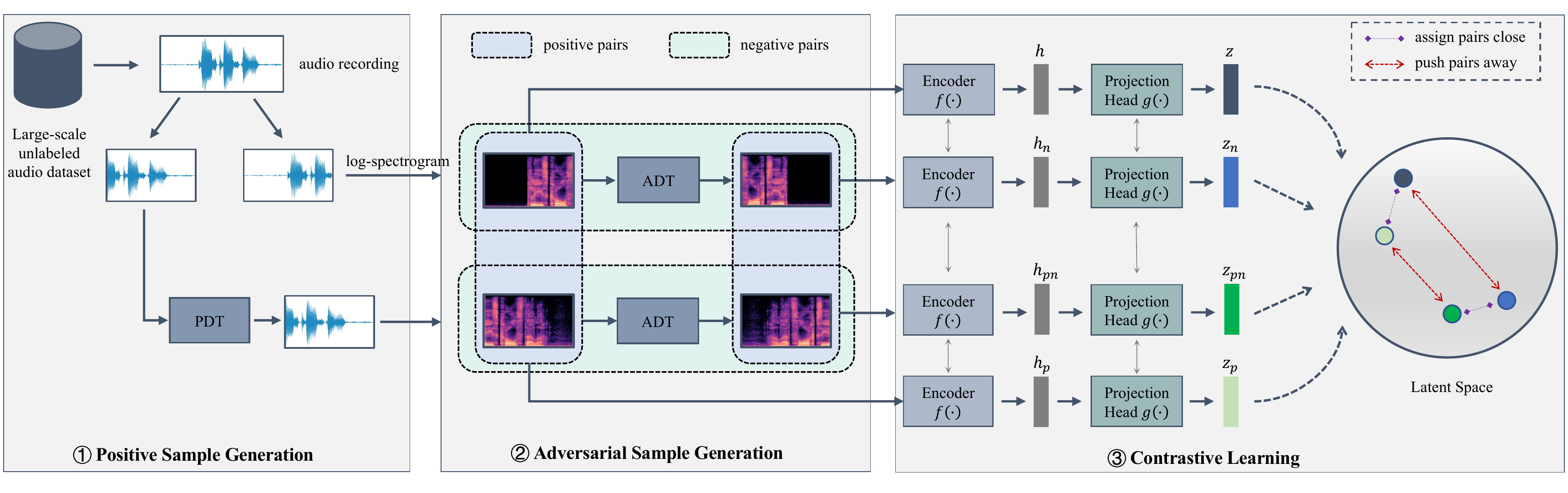}
  \caption{\textbf{Overall architecture of our pre-training framework}. PDT stands for positive data transformation and ADT stands for adversarial data transformation. Besides the common practice that pushing away spectrogram pairs sampled from different audio recordings, our framework asks the model to push away the adversarial pairs also.
  }
  \label{fig:framework}
  \vspace{-3mm}
\end{figure*}

\section{Approach}
%
We incorporate the proposed ASG mechanism into existing contrastive learning framework, \textit{i.e.}, SimCLR \cite{chen2020simple}, to further enhance the discriminative power of the learned audio representation, as shown in \cref{fig:framework}.
Unlike the previous methods that only generate positive pairs, we divide the data transformation strategies into two groups: \textit{(1) Positive sample generation}, which modifies the audio stream while retain its semantic. 
\textit{(2) Adversarial sample generation}, which employs positional transformations on spectrogram to generate negative samples.
Then, the NT-Xent Loss proposed in SimCLR is employed to optimize the network. The detailed descriptions of each part are given in the following sections.

\subsection{Positive Sample Generation}
Given an audio recording, we cut out two fix-length segments randomly and perform positive data transformation on one segment. They are considered as a positive pair, denoted as $x$ and $x_{p}$ respectively. 

\vspace{2pt}
\noindent\textbf{Positive data transformation}. We employs three transformations to generate the positive samples: 
\textit{(1) Room Impulse Response}, in which we convolve an audio stream with a set of room impulse responses to simulate the situations of playing audio in different rooms. 
%
%
\textit{(2) Random Tuning Volume}, which is realised by first randomly selecting a tuning factor in the volume tuning range and then multiplying the audio stream with the selected tuning factor.
\textit{(3) Adding White Noise}, which adds a uniform distribution with random intensity to audio stream. 
Some hyper-parameters used in positive sample generation are listed in Table \ref{tab:da_params}.

\begin{table}[!t]
  \caption{Parameters settings on positive data transformation.}
  \begin{tabular}{ccc}
    \toprule
    Transformations & Parameter  &   Value \\
    \midrule
    Room impulse response & Probability & 0.5 \\
    \hline
    \multirow{2}{*}{Random tuning volume} &
    Tuning range & [-10dB, 10dB] \\
    & Probability & 1.0 \\
    \hline
    \multirow{2}{*}{Adding White Noise} &
    Max intensity & 0.03 \\
    & Probability & 1.0 \\
    \bottomrule
  \end{tabular}
  \label{tab:da_params}
\end{table}


\subsection{Adversarial Sample Generation}
We convert the audio segments into spectrograms of shape $F \times T$ via Short-Time Fourier Transform (STFT), where $F$ stands for the number of frequency bins and $T$ stands for the number of time frames. 
We then transform the spectrograms to log-scale and feed them into adversarial data transformation to produce negative samples $x_{n}$ and $x_{pn}$ used for contrastive learning. 

\begin{figure}[!t]
  \centering
  \includegraphics[width=\linewidth]{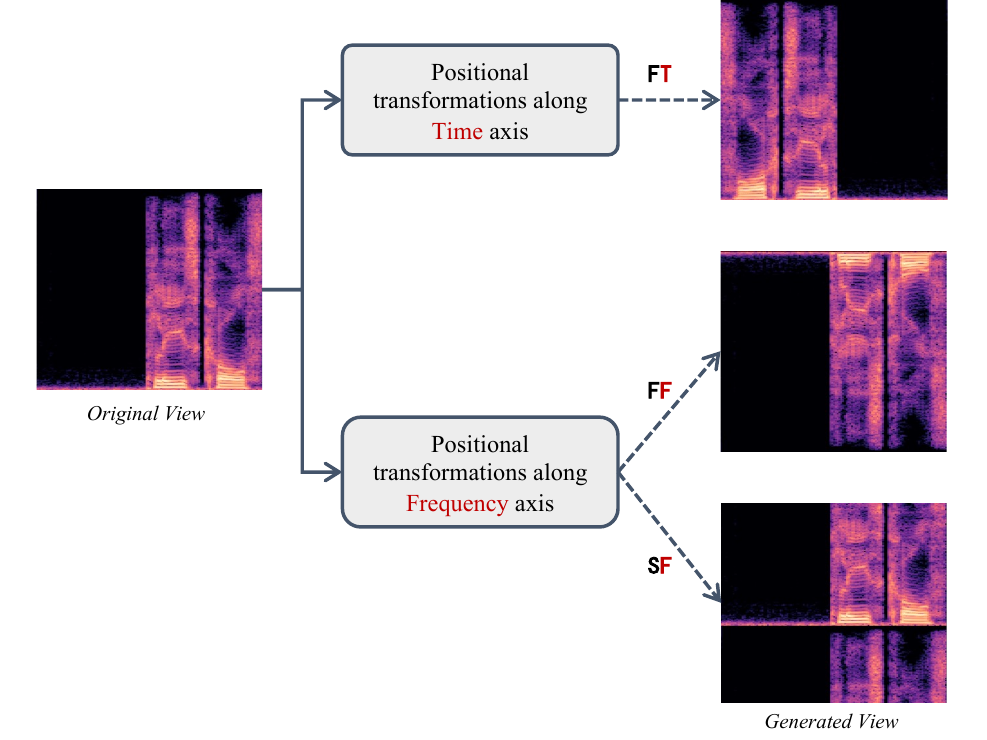}
  \caption{Three positional transformations employed in ASG. The generated adversarial sample share same appearance of visual patterns with the input, but at different direction or position.}
  \label{fig:nda}
  \vspace{-5mm}
\end{figure}

\vspace{2pt}
\noindent\textbf{Adversarial data transformation}. As shown in \cref{fig:nda}, the positional transformations can be divided into two types according to the central axis: 
(1) Positional transformations along time axis. We employ \emph{Flip along Time dimension} (FT) to emphasize the unidirectionality of time dimension of spectrogram. It challenges model to be sensitive to the direction of visual pattern in time dimension.
(2) Positional transformations along frequency axis. We employ \emph{Filp along Frequency dimension} (FF) and \emph{Scroll along Frequency dimension} (SF) transformations to emphasize the specificity of frequency distribution.
For SF, we introduce a hyper-parameter $\mathcal{L}_{ms}$ as the minimal length of scrolling along frequency dimension, due to that slight jittering is insufficient to change the semantics. 
When training, the scrolling length $\mathcal{L}_{\text{SF}}$ is randomly selected from $[\mathcal{L}_{ms}, F-\mathcal{L}_{ms}]$ at a uniform distribution, then we scroll the spectrogram along frequency dimension over $\mathcal{L}_{\text{SF}}$ length.

\subsection{Training}


Given a $N$-size batch of quadruples $(x, x_{p}, x_{n}, x_{pn})$, we reorganize them into $4N$ positive pairs, including $(x, x_{p})$, $(x_{p}, x)$, $(x_{n}, x_{pn})$, $(x_{pn}, x_{n})$. Other combinations are considered negative pairs.
We adopt ResNet-50 \cite{he2016deep} network as feature encoder $f$ to extract audio representation $\mathbf{h}=f\left(x\right) \in \mathbb{R}^{d_{h}}$. 
%
%
Then a multi-layer perceptron (MLP) is attached on the top of ResNet-50 as a projection head $g$, that maps $\mathbf{h}$ onto a space $\mathbf{z}=g \left( \mathbf{h} \right) \in \mathbb{R}^{d_{z}}$ for increasing flexibility, as described in SimCLR. The MLP is composed of a linear layer with output size of 4096, a batch normalization \cite{ioffe2015batch} layer, rectified linear units(ReLU) \cite{glorot2011deep} and a linear layer with output size of 256. Cosine similarity $s$ is measured, which is formulated as 

\begin{equation}
    s \left( \mathbf{z},\mathbf{z}^{\prime} \right) = \frac{\mathbf{z} ^\top}{||\mathbf{z}||} \cdot \frac{\mathbf{z}^{\prime}}{|| \mathbf{z}^{\prime}||} 
\end{equation}

The NT-Xent Loss proposed in SimCLR is utilized to optimize the parameters of encoder. For each positive pair index $\left( i, j \right)$ in batch input, the loss is calculated as 

\begin{equation}
    \mathscr{L}_{i,j} = -\log \frac{\exp \left( s \left(\mathbf{z}_{i}, \mathbf{z}_{j}\right) / \tau \right) }{\sum^{4N}_{k=1} \mathds{1}_{\left[ k \neq i \right]} \exp \left( s \left(\mathbf{z}_{i}, \mathbf{z}_{k}\right) / \tau \right) }
\end{equation}

where $\tau$ denotes as the temperature which is used to tune the scale of similarity.
\section{Experiments}

To verify the effectiveness of our ASG approach, we first learn the audio representation from the large-scale dataset (\textit{e.g.}, Audioset \cite{gemmeke2017audio} and VGGSound \cite{chen2020vggsound}) in a self-supervised manner, and then transfer the pre-trained representation to diverse downstream audio classification datasets. 
%
%
Following a generalized experimental setup in self-supervised learning, we measure the mean average precision (mAP) on Audioset and the top-1 accuracy on other datasets (\textit{e.g.}, LibriSpeech, voxCeleb1, VGGSound, \textit{etc.}) under two different evaluation settings: frozen evaluation and fine-tuned evaluation. The former only trains the linear classifier on top of a fixed encoder to fit downstream datasets, while the latter fine-tunes the whole framework in an end-to-end manner.

\subsection{Datasets}

\noindent\textbf{Pre-training dataset}.
%
%
Following \cite{saeed2021contrastive}, we use a large-scale audio dataset Audioset \cite{gemmeke2017audio} which is usually used for audio event recognition to pre-train the audio representation. Each sample in Audioset is a 10-second audio recording cut from a YouTube video. The dataset is splited into three parts: \textit{balanced train segments}, \textit{unbalanced train segments}, and \textit{eval segments}. Samples in balanced train segments and unbalanced train segments are used in our pre-training stage.
Since partial urls have been invalid, finally only 1,793,632 samples are used, which is less than the official number 1,963,807. 
In order to fully verify the effectiveness of our method, we employ another dataset VGGSound \cite{chen2020vggsound} on the ablation experiment. VGGSound is designed for audio event recognition task and multi-modal audio-visual analysis tasks. It is a large-scale audio dataset that consists of 200k videos for 309 audio classes, with a fixed duration of 10s for each sample.

\vspace{2pt}
\noindent\textbf{Downstream datasets}.
9 downstream datasets are employed to assess the performances of the pre-trained audio representation, including Audioset (AS), VGGSound (VS), librispeech (LS), Speech Commands V1 (SCV1), Speech Commands V2 (SCV2), NSynth (NS), voxCeleb1 (VC), MUSAN (MS) and ESC-50 (ESC). \textbf{Librispeech} \cite{panayotov2015librispeech} dataset is a corpus of approximately 1000 hours of English speech. The data is collected from audio of books read by 251 speakers and contains 33862 audio recordings in total. \textbf{Speech Commands V2} \cite{warden2018speech} dataset is a keyword spotting dataset with 35 spoken commands as classes. It contains 105829 examples in total with average duration of 1.0s. \textbf{Speech Commands V1} dataset is the same keyword spotting dataset with SCV2, but under different evaluation setting. It only contains ten basic commands that used in IoT or robotics applications, with the background noises as a new class. \textbf{NSynth} \cite{engel2017neural} dataset is used for musical instrument family classification with 11 family name classes. It contains 305979 samples with average duration of 4.0s. \textbf{voxCeleb1} \cite{nagrani2017voxceleb} dataset is used for speaker identification task, the same as Librispeech. It contains 153516 samples with average duration of 8.2s, collected from 1211 speakers. \textbf{MUSAN} \cite{snyder2015musan} dataset is used to differentiate audio samples across 3 classes (speech, music and noise). It contains only 2016 samples. \textbf{ESC-50} \cite{piczak2015esc} is an audio classifcation dataset consists of 2,000 5-second environmental audio recordings organized into 50 classes. 

\subsection{Implementation Details}

\noindent\textbf{Audio Pre-processing}.
Given an audio recording of arbitrary length, we decode it to a mono-channel digital audio stream, with sample rate of 16,000. Then we cut a fixed-length audio segment from it and convert the audio segment to a log-scaled spectrogram of shape $F \times T$ with window size of 32ms, hop size of 10ms and frequency range 50-8000Hz. $T$ stands for the number of frames in one spectrogram, corresponding to the duration of audio segment. $F$ stands for the number of frequency bins of spectrogram and equals to $L_{\text{FFT}} // 2 + 1$, where $L_{\text{FFT}}$ stands for the length of Fast Fourier Transform. In our experiments, $L_{\text{FFT}}=512$, corresponding to $F=257$.

\vspace{2pt}
\noindent\textbf{Pre-training Details}. 
We pre-train all representations using the LARS \cite{you2017large} optimizer with linear learning-rate scaling (\textit{i.e.}, \textit{LearningRate} $= 0.3 \times$ \textit{BatchSize} $/ 256$ ) 
and weight decay of $10^{-6}$. The number of frames $T$ is set to 300 (i.e., the shape of spectrogram input is $1 \times 257 \times 300$). We set the minimal length of scrolling $\mathcal{L}_{ms}=30$ and the temperature in contrastive loss $\tau=0.1$. We use linear warmup of learning rate and decay the learning rate with the cosine decay schedule without restarts. 
We train the network at batch size 512 for 50 epochs, with a linear warmup of learning rate for the first 2 epochs.
All the experiments are implemented in Python and run on Tesla-V100-32G GPU.

\vspace{2pt}
\noindent\textbf{Downstream Task Details}.
When training on downstream tasks, we split each audio into non-overlap segments and average their prediction scores as the final results. The number of frames $T$ in one segment is determined by the average duration of each dataset, following BYOL-A. We train the network at batch size 64 and a learning rate of $10^{-3}$ in both evaluation settings. For frozen evaluation, we train 100 epochs for each task and decay the learning rate by a factor of $0.1$ after 50 epochs. For fine-tuned evaluation, we train 50 epochs and decay the learning rate by a factor of $0.1$ after 25 epochs.

\subsection{Comparison with SOTAs} 
\begin{table*}[t]
  \centering\small
  \caption{ Results of comparison with state-of-the-art methods on 8 datasets. }
  \begin{tabular}{p{1.7cm}<{\centering}|p{2.2cm}<{\centering}p{1.1cm}<{\centering}p{1.1cm}<{\centering}p{1.1cm}<{\centering}p{1.1cm}<{\centering}p{1.1cm}<{\centering}p{1.1cm}<{\centering}p{1.1cm}<{\centering}p{1.1cm}<{\centering}p{1.1cm}<{\centering}}
    \hline
    \multirow{2}{*}{Evaluation}    & \multirow{2}{*}{Methods}    &   \multicolumn{8}{c}{Datasets}  \\
    &   & LS  &   SCV1   &   SCV2   &   MS   &   VC   &   NS  & AS & ESC \\
    \hline
    \hline
    \multirow{9}{*}{Frozen}
    & TRILL         &   -    &   74.0\%   &   -    &   -    &   17.7\%   &   -    & -  & -  \\
    & Bidir-CPC     &   -    &   -   &   92.4\%    &   -    &   40.7\%   &   58.4\%    & 28.7\% & 79.2\%  \\
    & Wav2Vec2.0    &   -    &   -   &   \textbf{94.8\%}    &   -    &   \textbf{41.6\%}   &   40.2\%    & 14.8\% & 69.2\%  \\
    \cline{2-10}
    & CBoW          &   99.0\%    &   -   &   30.0\%    &   98.0\%    &   -   &   33.5\%    & -  & - \\
    & SG            &   \textbf{100.0\%}    &   -   &   28.0\%    &   98.0\%    &   -   &   34.4\%    & - & -  \\
    & Temporal Gap  &   97.0\%    &   -   &   23.0\%    &   97.0\%    &   -   &   35.1\%    & - & -     \\    
    & COLA          &   \textbf{100.0\%}    &   71.7\%   &   62.4\%    &   99.1\%    &   29.9\%   &   63.4\%    & - & -  \\
    & BYOL-A        &   -    &   -   &   89.7\%    &   -    &   39.5\%   &   73.6\%    & 26.1\% & 82.3\%  \\
    
    & \textbf{Our ASG} & \textbf{100.0\%} & \textbf{87.9\%} & 83.2\% & \textbf{100.0\%} & 39.0\% & \textbf{76.0\%} & \textbf{35.3\%}  & \textbf{90.5\%}  \\
    \hline
    \hline
    \multirow{3}{*}{Fine-tuned} 
    & COLA          &   \textbf{100.0\%}    &   98.1\%   &   95.5\%    &   99.4\%    &   37.7\%   &   73.0\%    & - & - \\
    & SSAST         & -   &  -    &   \textbf{97.8\%}    &   -   &   57.1\%   &   -    & 29.0\% & 84.7\% \\
    & \textbf{Our ASG}    & \textbf{100.0\%} & \textbf{98.7\%} & 97.1\% & \textbf{100.0\%} & \textbf{74.7\%} & \textbf{80.9\%} & \textbf{42.4\%} & \textbf{93.2\%}  \\
    \hline
  \end{tabular}
  \label{tab:sota}
  \vspace{-3mm}
\end{table*}
We compare our ASG with state-of-the-art self-supervised pre-trained models. We evaluate transfer learning performance across 8 downstream tasks and report the overall results in Table \ref{tab:sota}. 
Specially, for frozen evaluation, we compare our ASG approach with 7 previous self-supervised methods, including Audio2Vec (CBoW and SG) \cite{tagliasacchi2020pre}, TemporalGap \cite{tagliasacchi2020pre}, COLA \cite{saeed2021contrastive}, BYOL-A \cite{niizumi2021byol}, TRILL \cite{shor2020towards}, Bidir-CPC \cite{wang2020contrastive} and Wav2Vec2.0 \cite{baevski2020wav2vec}.
We use the results reported on \cite{wang2022towards} for Bidir-CPC, Wav2Vec2.0 and BYOL-A and provide the results reported on \cite{saeed2021contrastive} for other methods. 
It is noteworthy that the Wav2Vec2.0 is pre-trained on LibriSpeech because it doesn't perform well when pre-trained with Audioset, as described in \cite{wang2022towards}. 
Among these methods, TRILL, Bidir-CPC and Wav2Vec2.0 are proposed to learn representation for speech task while others tend to learn general-purpose audio representation. Bidir-CPC and Wav2Vec2.0 operate on raw audio sequence directly, which preserves the temporal structure of audio so that they are suitable for speech recognition task. 
From table \ref{tab:sota} we can see Wav2Vec2.0 show big superiority on speech task, surpassing our ASG by 11.6\% and 2.6\% on SCV2 and VC respectively. However, it perform much worse on other tasks. Especially on NS, Wav2Vec2.0 achieves 35.8\% performance drop compared with our ASG, which indicates that it is not discriminative enough for non-speech audio classification. 
Considering only spectrogram-based methods, our ASG achieves best or comparable performance on most benchmarks, demonstrating its effectiveness on audio representation learning. 
Specifically, our ASG achieves gains of \textbf{13.9\%} / \textbf{0.9\%} / \textbf{2.4\%} / \textbf{9.2\%} / \textbf{8.2\%} on SCV1, MS, NS, AS and ESC respectively.
However, we can observe a 6.5\% performance drop when compared with BYOL-A on SCV2. A reason for this observation is BYOL-A augments audio with random resize cropping, making learned representation robust to flexible temporal scales, which plays an important role on speech recognition task but is insignificant for acoustic scene classification.
For fine-tuned evaluation, we compare our ASG method with two state-of-the-art methods, \textit{i.e.}, COLA and SSAST \cite{gong2022ssast}. 
Our ASG achieves \textbf{0.6\%} / \textbf{0.6\%} / \textbf{17.6\%} / \textbf{7.9\%} / \textbf{13.4\%} / \textbf{8.5\%} improvements on SCV1, MS, VC, NS, AS and ESC respectively, which demonstrates the superiority of our method.

\subsection{Comparison with Supervised Pre-trained Models}

\begin{table}[t]
    \centering\small
    \caption{ Comparison with supervised pre-trained model on 8 datasets. Sup. denotes the supervised pre-trained model.}
    \setlength{\tabcolsep}{9pt}{
    \begin{tabular}{ccccc}
    \hline
    \multirow{2}{*}{Datasets} & \multicolumn{2}{c}{Frozen} & \multicolumn{2}{c}{Fine-tuned} \\
    & Sup. & ASG   & Sup. & ASG \\
    \hline 
    \hline
    LS   &   99.0\%     &   \textbf{100.0\% }    &   100.0\%     &   \textbf{100.0\%} \\
    SCV1 &   86.6\%     &   \textbf{87.9\%}      &   97.8\%      &   \textbf{98.7\%} \\
    SCV2 &   82.1\%     &   \textbf{83.2\%}      &   96.4\%      &   \textbf{97.1\%} \\
    MS   &   100.0\%    &   \textbf{100.0\%}     &   100.0\%     &   \textbf{100.0\%} \\
    VC   &   20.4\%     &   \textbf{39.0\%}      &   65.2\%      &   \textbf{74.7\%} \\
    NS   &   71.9\%     &   \textbf{76.0\%}      &   78.5\%      &   \textbf{80.9\%} \\
    AS   &   -          &   -                    &   39.9\%      &   \textbf{42.4\%} \\
    ESC  &   90.0\%     &   \textbf{90.5\%}      &   91.1\%      &   \textbf{93.2\%} \\
    \hline 
    \end{tabular}
    }
    \label{tab:supervise}
    \vspace{-3mm}
\end{table}
We then show the comparison results of self-supervised and supervised manners to further evaluate the transfer learning performance of our proposed ASG. 
For supervised pre-training, we follow PANNs \cite{kong2020panns} to train CNN on Audioset with a multi-label classification loss at batch size 1024 for 52,560 steps. 
We use LARS optimizer with a learning rate of 0.4 and decay the learning rate with the cosine decay schedule without restarts. Class-balanced data sampling strategy, mixup and SpecAugement \cite{park2019specaugment} methods are employed to mitigate over-fitting. 
For fair compaison, resnet50 is adopted as backbone and log-spectrogram is used as network input.
We assess their performances on 8 downstream datasets and report the results in Table \ref{tab:supervise}. 
Our ASG model achieves superior or competitive results on all datasets. Especially on voxCeleb1, ASG achieves \textbf{18.6\%} / \textbf{9.5\%} improvements compared with supervised pre-trained resnet50, under frozen and fine-tuned evaluation respectively. These results indicate that our self-supervised pre-trained representation is robust enough and can be used as initialization for various downstream tasks.

\subsection{Ablation Studies}

\begin{table}[!t]
  \centering\small
  \caption{Ablation study on adversarial data transformations.}
  \setlength{\tabcolsep}{7pt}{
  \begin{tabular}{lccccc}
  \hline
  \multirow{2}{*}{Datasets} &   \multicolumn{5}{c}{Adversarial Location Transformation} \\
  &    NONE    &   ST &   FT  &   SF  &   FF   \\
  \hline
  \hline
  LS   & 99.1\%  & 99.5\%  & 99.6\%  & 99.6\% & \textbf{99.8\%} \\
  SCV1 & 84.7\%  & 84.4\%  & \textbf{86.8\%}  & 85.1\% & 85.4\% \\
  SCV2 & 78.9\%  & 79.0\%  & \textbf{81.8\%}  & 79.1\% & 80.3\% \\
  VC   & 21.8\%  & 24.9\%  & 25.3\% & 25.2\%  & \textbf{27.3\%}  \\
  NS   & 65.3\%  & 64.9\%  & \textbf{67.7\%}  & 67.6\% & 67.6\%  \\
  \hline
  Average & 70.0\%  & 70.5\%  & \textbf{72.2\%} & 71.3\% & 72.1\%  \\
  \hline
  \end{tabular}
  }
  \label{tab:asg}
\end{table}
\noindent\textbf{Adversarial data transformation}.
We conduct experiments to clarify the contribution of each proposed adversarial data transformation. The results are shown in Table \ref{tab:asg}. Three adversarial transformations are proposed in this paper, namely Flip along Time dimension (FT), Flip along Frequency dimension (FF), Scroll along Frequency dimension (SF) respectively. 
In this section, the three transformations are separately applied to generate adversarial samples, instead of a random selection from all three transformations in ASG, to evaluate the effectiveness of each adversarial transformation. 
The result of the model pre-trained without using ASG is reported as baseline, denoted as NONE.
Besides, a model that trained with Scroll along Time axis (ST) is also evaluated as a comparison.

We pre-train all representations under the same setup and evaluate them under frozen evaluation on 5 downstream datasets. From Table \ref{tab:asg}, we can observe that:
(1) All three proposed adversarial transformations(\textit{i.e.}, FT, SF and FF), perform better than the baseline(\textit{i.e.}, NONE). Specifically, FT, SF, FF outperforms NONE by \textbf{2.2\%, 1.3\%, 2.1\%}, respectively.
It verifies that audio semantic is really sensitive to positional change, and each proposed adversarial transformation plays a positive role on audio contrastive learning.
(2) FT outperforms ST and achieves an \textbf{1.7\%} improvement on average, which suggests it is the direction of visual pattern rather than the absolute position in time dimension plays a decisive role in audio semantics.
(3) The improvement of ST is negligible compared with the other three transformations and it even makes negative effect on some tasks, such as SCV1 and NS. So it is not suitable on general-purpose audio representation learning.

\vspace{2pt}

\begin{table}[!t]
  \centering\small
  \caption{Performance with mel-spectrogram as input. }
  \setlength{\tabcolsep}{8pt}{
  \begin{tabular}{lcccc}
    \hline
    \multirow{2}{*}{Datasets}    &    \multicolumn{2}{c}{Frozen}    &   \multicolumn{2}{c}{Fine-tuned}  \\
    &    w/o ASG &   w/ ASG   &    w/o ASG &   w/ ASG  \\
    \hline
    \hline
    LS      & 99.9\% & \textbf{100.0\%} & 100.0\% & \textbf{100.0\%} \\
    SCV1    & 85.7\% & \textbf{87.2\%} & 97.2\% & \textbf{97.4\%} \\
    SCV2    & 79.1\% & \textbf{79.3\%} & 97.0\% & \textbf{97.1\%} \\
    VC      & 34.2\% & \textbf{35.3\%} & 51.4\% & \textbf{53.8\%} \\
    NS      & 66.3\% & \textbf{69.1\%} & 73.2\% & \textbf{75.2\%} \\
    \hline
    Average & 73.0\% & \textbf{74.2\%} & 83.8\% & \textbf{84.7\%}   \\
    \hline
  \end{tabular}
  }
  \label{tab:mel}
  \vspace{-2mm}
\end{table}
\noindent\textbf{Type of spectrogram}.
We explore the effectiveness of ASG with mel-spectrogram as input and report the experimental results in Table \ref{tab:mel}. From the table we can observe that model pre-trained with ASG performs better than model pre-trained without ASG on all experiment settings, achieving \textbf{1.2\%} / \textbf{0.9\%} performance gains on average, under frozen and fine-tuned evaluation respectively. This observation is consistent with experiments using spectrogram as input, and further demonstrates that audio semantic is sensitive to the positional transformation of its spectrogram.


\vspace{2pt}
\begin{table}[!t]
  \centering\small
  \caption{Ablation study on pre-train data. }
  \setlength{\tabcolsep}{11pt}{
  \begin{tabular}{clcc}
    \hline
    \multirow{2}{*}{Pretrain Data}  & \multirow{2}{*}{Method}   &  \multicolumn{2}{c}{Evaluation Setting}   \\
    &   &   Frozen  &   Fine-tuned \\
    \hline
    \hline
    \multirow{3}{*}{Audioset}   &   Sup.        & -         &   39.9\%   \\
                                &   w/o ASG     & 31.8\%    &   41.1\%   \\
                                &   w/ ASG      & \textbf{35.3\%}    &   \textbf{42.4\%}   \\
    \hline
    \hline
    \multirow{3}{*}{VGGSound}   &   Sup.        & -         &   50.5\%   \\
                                &   w/o ASG     & 41.6\%    &   51.5\%   \\
                                &   w/ ASG      & \textbf{44.7\%}    &   \textbf{53.0\%}   \\
    \hline
  \end{tabular}
  }
  \label{tab:data}
  \vspace{-5mm}
\end{table}
\noindent\textbf{Pre-training data}.
Finally, to evaluate the generalization ability of our proposed ASG, we compare the performances of supervised pre-trained model (denoted as Sup.), self-supervised pre-trained model without ASG (denoted as w/o ASG) and self-supervised pre-trained model with ASG (denoted as w/ ASG), under two large-scale pre-train dataset, \textit{i.e.}, Audioset and VGGSound. Table \ref{tab:data} shows that models trained with ASG achieves \textbf{3.5\%} / \textbf{1.3\%} improvements on Audioset and \textbf{3.1\%} / \textbf{1.5\%} improvements on VGGSound, under frozen and fine-tuned evaluation respectively. Besides, our ASG outperforms the model trained from scratch by \textbf{2.5\%} / \textbf{2.5\%} when fine-tuning on Audioset and VGGSound respectively, which demonstrates the effectiveness of ASG.

\section{Conclusion}

We discover the sensitivity of audio spectrogram to positional transformations, which is important to audio analysis. 
Therefore, we introduce an Adversarial Sample Generation (ASG) mechanism, to enhance the discriminative power of learned audio representation. 
It first constructs the negative pair by flipping or translating spectrogram along time or frequency axis, then enforces the instances of negative pair to be pushed away in the latent space, driving CNN more sensitive to the positional information in spectrogram. 
Our experiment results demonstrate the effectiveness of ASG and extensive ablation studies are conducted to clarify the contribution of each adversarial data transformation. 
%

{\small
\bibliographystyle{abbrv}
\bibliography{ref}
}

\end{document}